\documentclass[aps,showpacs, nofootinbib]{revtex4}

\newcommand{\be}{\begin{equation}}
\newcommand{\ee}{\end{equation}}
\newcommand{\ben}{\begin{eqnarray}}
\newcommand{\een}{\end{eqnarray}}

\newcommand{\cO}{{\cal O}}

\newcommand{\p}{\partial}
\newcommand{\na}{\nabla}

\newcommand{\tpsi}{\tilde \psi}
\newcommand{\tim}{{\tilde \mu}}
\newcommand{\tom}{{\tilde \omega}}

\newcommand{\ep}{\epsilon}


\begin{document}

\title{Decay of Massive Scalar Hair in the Background of Dilaton Gravity Black Hole}

\author{Marek Rogatko}
\affiliation{Institute of Physics \protect \\
Maria Curie-Sklodowska University \protect \\
20-031 Lublin, pl.~Marii Curie-Sklodowskiej 1, Poland \protect \\
rogat@tytan.umcs.lublin.pl \protect \\
rogat@kft.umcs.lublin.pl}

\date{\today}

\begin{abstract}
We investigate analytically both the intermediate and late-time behaviour of the massive 
scalar field in the background of static spherically symmetric black hole solution in dilaton 
gravity with arbitrary coupling constant. The intermediate asymptotic behaviour of scalar 
field depends on the field's parameter mass as well as the multiple number $l$.
On its turn, the late-time behaviour has the power law decay rate independent on coupling constant
in the theory under consideration. 
\end{abstract}

\maketitle

\section{Introduction}

Late-time behaviour of various fields in the spacetime of a collapsing
body plays an important role
in black hole's physics. It happens that regardless of details of the collapse or the structure and properties
of the collapsing body the resultant black hole can be described only by few parameters
such as mass, charge and angular momentum, {\it black holes have no hair}. 
On its own,
it is interesting to investigate how these hair loss proceed dynamically. 
\par
For the first time the neutral external perturbations were studied by Price in \cite{pri72}. It was 
shown that the late-time behavior is dominated by the factor $t^{-(2l + 3)}$, for each 
multipole moment $l$. In Ref.\cite{gun94} it was revealed that the decay-rate 
along null infinity and along the
future event horizon was governed
by the power laws $u^{-(l + 2)}$ and $v^{-(l + 3)}$, where $u$ and $v$ were
the outgoing Eddington-Finkelstein (ED) and ingoing ED coordinates.
On the other hand, Ref. \cite{bic72} was devoted to the scalar perturbations on Reissner-Nordstr\"om (RN)
background for the case when  $\mid Q \mid < M$ has the following dependence on time $t^{-(2l + 2)}$,
while for $\mid Q \mid = M$ the late-time behavior at fixed $r$ is governed by
$t^{-(l + 2)}$. 
\par
It turns out that a charged hair
decayed slower than a neutral one \cite{pir1}-\cite{pir3}, while
the late-time tails in gravitational collapse of a self-interacting (SI)
fields in the background of Schwarzschild solution was reported by Burko \cite{bur97}
and in
RN solution at intermediate late-time was considered
in Ref.\cite{hod98}. At intermediate late-time for small mass $m$ the decay was dominated
by the oscillatory inverse power tails $t^{-(l +3/2)} \sin (m t)$. This analytic prediction
was verified at intermediate times, where $mM \ll mt \ll 1/(mM)^2$. 
It was found analytically 
\cite{ja} that for
a nearly extreme RN spacetime the inverse power law behavior of the dominant asymptotic tail is of the form
$t^{-5/6} \sin (m t)$. Just it is independent of $l$. The asymptotic tail behaviour 
of SI scalar field
was also studied in Schwarzschild spacetime \cite{ja1}. The oscillatory
tail of scalar field has the decay rate of $t^{-5/6}$ at                                   
asymptotically late time.
The power-law tails in the evolution of a charged massless scalar field around a fixed
background of dilaton black hole was studied in Ref.\cite{mod01a}, where
the inverse power-law relaxation of the fields at future timelike infinity, future null infinity and
along the outer horizon of the considered black hole was found.
The case of a massive 
scalar field was elaborated in \cite{mod01b} and it was envisaged numerically that at very late times
the oscillatory tail decay is of the form $t^{-5/6}$.
\par
In addition, the late-time behaviour of massive Dirac fields were studied
in the spacetime of Schwarzschild black hole \cite{jin04}. It was found that the asymptotic behaviour of these fields
is dominated by a decaying tail without any oscillation. The dumping exponent was independent on
the multiple number of the wave mode and on the mass of the Dirac field. On the other hand,
the decay of the massive Dirac fields was slower comparing the decay of massive scalar field.
The above analysis was supplemented by the studies of charged massive Dirac fields in the spacetime of
RN black hole \cite{jin05}. 
The case of the decay of the fields in the stationary axisymmetric black hole background was studied numerically
in Ref.\cite{bur04} and it was found that in the case of Kerr black hole that the oscillatory inverse-power law of 
the dominant asymptotic tail behaviour is approximately depicted by the relation $t^{-5/6}\sin(mt)$.
In Ref.\cite{xhe06} both the intermediate late-time tail and the asymptotic behaviour of the charged massive
Dirac fields in the background of Kerr-Newmann black hole was investigated. It was 
demonstrated that 
the intermediate late-time behaviour of the fields under consideration is dominated by an
inverse power-law decaying tail without any oscillation.
\par
The late-time behaviour of massive vector field obeying the Proca equation of motion in the background
of Schwarzschild black hole was studied in \cite{kon06}. It was revealed that 
at intermediate late times,
three functions characterizing
the field have different decay law depending on the multiple number $l$. On the contrary,
the late-time behaviour is independent on $l$, i.e., the late-time decay law is
proportional to $t^{-5/6}\sin(mt)$.
\par
As far as the  $n$-dimensional static black holes is concerned, the
{\it no-hair} theorem for them
is quite well established \cite{unn}. The  mechanism of decaying black hole hair in higher dimensional static
black hole case concerning the evolution of massless scalar field in the $n$-dimensional Schwarzshild
spacetime was determined in
Ref.\cite{car03}. It was found that for odd dimensional spacetime the field  decay had a
power falloff like $t^{-(2l + n - 2)}$, where $n$ is the dimension of the spacetime.
This tail was independent of the presence of the black hole.
For even dimensions the late-time behaviour is also in the power law form but in this case it is due to
the presence of black hole $t^{-(2l + 3n - 8)}$.
The late-time tails of massive scalar fields in the spacetime of $n$-dimensional 
static charged black hole was elaborated in Ref.\cite{mod05} 
and it was found that the intermediate asymptotic behaviour of massive scalar field
had the form $t^{-(l + n/2 - 1/2)}$. This pattern of decay was checked numerically for $n = 5, 6$.
The massive scalar field quasi-normal 
modes in higher dimensional black hole were considered in \cite{zhi06}. Among all it was envisaged
a qualitatively different dependence of the fundamental modes on the field mass for $n \ge 6$.
\par
In our work we shall discuss analytically the intermediate and late-time behaviour of
massive scalar field in the spacetime of black hole in dilaton gravity with arbitrary coupling constant.
We confirmed the numerical result presented in Ref.\cite{mod01b}. Namely, we revealed that the intermediate asymptotic 
behaviour is not the final pattern of the decay of the massive scalar hair. At late times the inverse power law
of the decay is relevant, which is independent on field's parameters and the coupling constant of the theory.
In Sec.II we gave the analytic arguments concerning
the decay of scalar massive hair in the background of the considered black hole. 
We conclude our investigations and give some remarks concerning the extremal dilaton black holes in Sec.III.

\section{Green's function analysis }
The action for the dilaton gravity with arbitrary coupling constant is subject to the form
\be
S = \int d^4x \sqrt{-g} \bigg[ R - 2 \na^{\mu} \phi \na_{\mu} \phi - e^{- 2 \alpha \phi} 
F_{\mu \nu}  F^{\mu \nu} \bigg],
\ee
where $\phi$ is the dilaton field, $\alpha$ coupling constant while $F_{\mu \nu} = 2 \na_{[\mu} A_{\nu]}$
is the strength of $U(1)$ gauge field.\\
The static spherically symmetric solution of the Eqs. of motion for the underlying theory
may be written as \cite{gar91}
\be
ds^2 = - \bigg( 1 - {r_{+} \over r} \bigg)\bigg( 1 - {r_{-} \over r} \bigg)^{{1 - \alpha^2} \over 1 + \alpha^2} dt^2
+ {dr^2 \over \bigg( 1 - {r_{+} \over r} \bigg)\bigg( 1 - {r_{-} \over r} \bigg)^{{1 - \alpha^2} \over 1 + \alpha^2}}
+ R^2(r) d\Omega^2,
\label{dila}
\ee
where $R^2(r) = r^2 \bigg( 1 - {r_{-} \over r} \bigg)^{2 \alpha^2 \over 1 + \alpha^2}$, while $r_{+}$ and $r_{-}$
are related to the mass $M$ and the electric charge $Q$ of the black hole
\be
e^{2 \alpha \phi} = \bigg( 1 - {r_{-} \over r_{+}} \bigg)^{2 \alpha^2 \over 1 + \alpha^2}, \qquad
2 M = r_{+} + {1 - \alpha^2 \over 1 + \alpha^2} r_{-}, \qquad
Q^2 = {r_{-}~r_{+} \over 1 + \alpha^2}.
\label{ext}
\ee
The metric is asymptotically flat in the sense that
the spacetime  
contains a data set
$(\Sigma_{end}, g_{ij}, K_{ij})$ with gauge fields such that 
$\Sigma_{end}$ is diffeomorphic to ${\bf R}^3$ minus a ball and the 
following asymptotic conditions are fulfilled:
\ben
\vert g_{ij}  - \delta_{ij} \vert + r \vert \p_{a}g_{ij} \vert
+ ... + r^k \vert \p_{a_{1}...a_{k}}g_{ij} \vert +
r \vert K_{ij} \vert + ... + r^k \vert \p_{a_{1}...a_{k}}K_{ij} \vert
\le {\cal O}\bigg( {1\over r} \bigg), \\
\vert F_{\alpha \beta} \vert + r \vert \p_{a} F_{\alpha \beta} \vert
+ ... + r^k \vert \p_{a_{1}...a_{k}}F_{\alpha \beta} \vert
\le {\cal O}\bigg( {1 \over r^2} \bigg),\\
\phi = \phi_{0} + {\cal O}\bigg( {1\over r} \bigg),
\een
where $K_{ij}$ is the exterior curvature, $\phi_{0}$ is a constant value of the scalar field.\\
The massive scalar field $\tpsi$ satisfies the following Eq. of motion:
\be
\na^{i}\na_{i}\tpsi - m^2 \tpsi = 0.
\ee
Next, we define the tortoise coordinates $y$ as
\be
dy = {dr \over 
\bigg( 1 - {r_{+} \over r} \bigg)\bigg( 1 - {r_{-} \over r} \bigg)^{{1 - \alpha^2} \over 1 + \alpha^2}}.
\ee
Thus, the metric (\ref{dila}) can be rewritten in the form as
\be
ds^2 = \bigg( 1 - {r_{+} \over r} \bigg)\bigg( 1 - {r_{-} \over r} \bigg)^{{1 - \alpha^2} \over 1 + \alpha^2}
\bigg( -dt^2 + dy^2 \bigg) + R^2(r) d\Omega^2.
\ee
We resolve the field into spherical harmonics, which leads to the relation
\be
\tpsi = \sum_{l,m} {1 \over R(r)} \psi_{m}^{l}(t, r) 
Y_{l}^{m}(\theta, \phi).
\ee
where $Y_{l}^{m}$ is a scalar spherical harmonics on the unit two-sphere.
As a consequence of the above one gets the following equations of motion for each multipole moment
\be
\psi_{,tt} - \psi_{,yy} + V \psi = 0.
\label{mo}
\ee
The effective potential $V$ implies the following relation:
\be
V = \bigg( 1 - {r_{+} \over r} \bigg)\bigg( 1 - {r_{-} \over r} \bigg)^{{1 - \alpha^2} \over 1 + \alpha^2} 
\bigg[ 
{1 \over R(r)} {d \over dr}\bigg(
\bigg( 1 - {r_{+} \over r} \bigg)\bigg( 1 - {r_{-} \over r} \bigg)^{{1 - \alpha^2} \over 1 + \alpha^2} {dR(r)\over dr}
\bigg) + {l(l + 1) \over R^2(r)} + m^2 \bigg].
\ee
In order to analyze the time evolution of a massive field in the background of
dilaton black hole we shall use the spectral decomposition method.
In Refs.\cite{hod98},\cite{lea86} it was shown that the asymptotic tail is connected with the
existence of a branch cut situated along the interval $-m \le \omega \le m$.
An oscillatory inverse power-law behaviour of the self-interacting scalar field arises
from the integral of Green function $\tilde G(y, y';\omega)$ around branch cut.
In what follows we shall denote it by $G_{c}$ and our main aim will be to find the 
analytic form of this integral.\\
Namely,
the time evolution of massive scalar field may be written in the following form:
\be
\psi(y, t) = \int dy' \bigg[ G(y, y';t) \psi_{t}(y', 0) +
G_{t}(y, y';t) \psi(y', 0) \bigg],
\ee
for $t > 0$, where   the Green's function  $ G(y, y';t)$ is given by the relation
\be
\bigg[ {\p^2 \over \p t^2} - {\p^2 \over \p y^2 } + V \bigg]
G(y, y';t)
= \delta(t) \delta(y - y').
\label{green}
\ee
In what follows, 
our main task will be to find the black hole Green function.
In the first step we reduce equation
(\ref{green}) to an ordinary differential equation.
To do it one can use the Fourier transform \cite{lea86}
$\tilde  
G(y, y';\omega) = \int_{0^{-}}^{\infty} dt~ G(y, y';t) e^{i \omega t}$.
This Fourier's transform is well defined for $Im~ \omega \ge 0$, while the 
corresponding inverse transform yields
\be
G(y, y';t) = {1 \over 2 \pi} \int_{- \infty + i \ep}^{\infty + i \ep}
d \omega~
\tilde G(y, y';\omega) e^{- i \omega t},
\ee
for some positive number $\ep$.
The Fourier's component of the Green's function $\tilde  G(y, y';\omega)$
can be written in terms of two linearly independent solutions for
homogeneous equation as
\be
\bigg(
{d^2 \over dy^2} + \omega^2 - V \bigg) \psi_{i} = 0, \qquad i = 1, 2.
\label{wav}
\ee
The boundary conditions for $\psi_{i}$ are described by purely ingoing waves
crossing the outer horizon $H_{+}$ of the 
charged dilaton black hole
$\psi_{1} \simeq e^{- i \omega y}$ as $y \rightarrow  - \infty$ while
$\psi_{2}$ should be damped exponentially at $i_{+}$, namely
$\psi_{2} \simeq e^{- \sqrt{m^2 - \omega^2}y}$ at $y \rightarrow \infty$.

To proceed further, we change the variables
\be
\psi_{i} = {\xi \over \bigg( 1 - {r_{+} \over r} \bigg)^{1/2}
\bigg( 1 - {r_{-} \over r} \bigg)^{{1 - \alpha^2} \over 2(1 + \alpha^2)}},
\ee
where $i = 1,2$. 
Let us assume that the observer and the initial data are situated far away from the considered
black hole. We expand Eq.(\ref{wav}) as a power  series of $ r_{\pm}/r$ neglecting terms of order
$\cO ((r_{\pm}/r)^2)$ and higher. It leads us to the following expression:
\be
{d^2 \over dr^2} \xi + \bigg[
\omega^2 - m^2 + {(2 \omega^2 - m^2)( r_{+} + \alpha_{1} r_{-}) \over r} -{l(l + 1) \over r^2}
\bigg] \xi = 0,
\ee
where $\alpha_{1} = {{1 - \alpha^2} \over 1 + \alpha^2}$.\\ 
It turned out that the above equation can be solved in terms of Whittaker's functions \cite{abr70}, namely
the two basic solutions are needed to construct the Green function, with the condition that
$\mid \omega \mid \ge m$. Consequently, it implies the result as follows:
\be
\tpsi_{1} = M_{\kappa, \tim}(2 \tom r),  \qquad 
\tpsi_{2} = W_{\kappa, \tim}(2 \tom r),
\ee
where we have denoted
\be
\tim = \sqrt{ 1/4 + l(l + 1)}, \qquad    \kappa =  ( r_{+} + \alpha_{1} r_{-})\bigg(
{m^2 \over 2 \tom} - \tom \bigg), \qquad        
\tom^2 = m^2 - \omega^2.
\ee
By virtue of the above relations the spectral Green function yields
\ben
G_{c}(x,y;t) &=& {1 \over 2 \pi} \int_{-m}^{m}dw
\bigg[ {\tpsi_{1}(x, \tom e^{\pi i})~\tpsi_{2}(y,\tom e^{\pi i}) \over W(\tom e^{\pi i})}
- {\tpsi_{1}(x, \tom )~\tpsi_{2}(y,\tom ) \over W(\tom )} 
\bigg] ~e^{-i w t} \\ \nonumber
&=& {1 \over 2 \pi} \int_{-m}^{m} dw f(\tom)~e^{-i w t},
\een 
where by we denoted Wronskian $W(\tom)$.\\ 
First we discuss 
the intermediate asymptotic behaviour of the massive scalar field, i.e., when the range of parameters are 
$M \ll  r \ll t \ll M/(m M)^2$.
The intermediate asymptotic contribution to the Green function integral gives the frequency equal to 
$\tom = {\cO (\sqrt{m/t})}$. It implies that $\kappa \ll 1$. One should have in mind that $\kappa$ 
stems from the $1/r$ term in the massive scalar field equation of motion. Thus it depicts
the effect of backscattering off the spacetime curvature and in the case under consideration
the backscattering is negligible. In the case of intermediate asymptotic behaviour
one finally gets
\be
f(\tom) = {2^{2 \tim -1} \Gamma(-2\tim)~\Gamma({1 \over 2} + \tim) \over
\tim \Gamma(2 \tim)~\Gamma({1 \over 2} - \tim)} \bigg[
1 + e^{(2 \tim + 1) \pi i} \bigg]
(r r')^{{1 \over 2} + \tim} \tom^{2 \tim},
\ee
where we have used the fact that $\tom r \ll 1$ and the form of $f(\tom)$
can be approximated by means of the fact that $M(a, b, z) = 1$ as $z$ tends to zero.
The resulting Green function can be written as
\be
G_{c}(r,r';t) = {2^{3 \tim - {3\over2}} \over \tim \sqrt{\pi}}
{\Gamma(-2\tim)~\Gamma({1 \over 2} + \tim) \Gamma(\tim +1 ) \over
\tim \Gamma(2 \tim)~\Gamma({1 \over 2} - \tim)}
\bigg( 1 + e^{(2 \tim + 1) \pi i} \bigg)~(r r')^{{1 \over 2} + \tim} 
~\bigg( {m\over t} \bigg)^{{1 \over 2} + \tim}~J_{{1 \over 2} + \tim}(mt).
\ee  
In the limit when $t \gg 1/m$ one arrives at the conclusion that the spectral Green function provides
\be
G_{c}(r',r;t) = {2^{3 \tim - 1} \over \tim \sqrt{\pi}}
{\Gamma(-2\tim)~\Gamma({1 \over 2} + \tim) \Gamma(\tim +1 ) \over
\tim \Gamma(2 \tim)~\Gamma({1 \over 2} - \tim)}
\bigg( 1 + e^{(2 \tim + 1) \pi i} \bigg)~(r r')^{{1 \over 2} + \tim} 
~m^{\tim}~ t^{- 1 - \tim}~\cos(mt - {\pi \over 2}(\tim + 1)).
\label{gfim}
\ee  
Eq.(\ref{gfim}) depicts the oscillatory inverse power-law behaviour. In our case the intermediate
times of the power-law tail depends only on $\tim$ which in turn is a function of the angular
momentum parameter $l$.
\par
However, the different pattern of decay is expected when  $\kappa \gg 1$, for the late-time
behaviour, when the backscattering off the curvature is important.
In this case we take into account the limit \cite{abr70}
\be
M_{\kappa, \tim}(2 \tom r) \approx \Gamma (1 + 2 \tim)~(2 \tom r)^{1 \over 2}
\kappa^{- \tim}~J_{2 \tim} (\sqrt{8 \kappa \tom r}).
\ee
Consequently, $f(\tom)$ yields
\ben \label{fer}
f(\tom) &=& {\Gamma(1 + 2\tim)~\Gamma(1 - 2\tim) \over 2 \tim}~(r r')^{1 \over 2}
\bigg[ J_{2 \tim} (\sqrt{8 \kappa \tom r})~J_{- 2 \tim} (\sqrt{8 \kappa \tom r'})
- I_{2 \tim} (\sqrt{8 \kappa \tom r})~I_{- 2 \tim} (\sqrt{8 \kappa \tom r'}) \bigg] \\ \nonumber
&+&
{(\Gamma(1 + 2\tim))^2~\Gamma(- 2\tim)~\Gamma( {1 \over 2} + \tim - \kappa)
 \over 2 \tim ~\Gamma(2 \tim)~\Gamma({1 \over 2} - \tim - \kappa) }~(r r')^{1 \over 2}
~\kappa^{- 2 \tim}
\bigg[
J_{2 \tim} (\sqrt{8 \kappa \tom r})~J_{2 \tim} (\sqrt{8 \kappa \tom r'})
\\ \nonumber
&+& e^{(2 \tim + 1)}
I_{2 \tim} (\sqrt{8 \kappa \tom r})~I_{2 \tim} (\sqrt{8 \kappa \tom r'}) 
\bigg].
\een
It can be noticed that the first part of the above Eq.(\ref{fer}) the late time tail is proportional to $t^{-1}$. 
Just we calculate
the second term of the right-hand side of Eq.(\ref{fer}). For the case when 
$\kappa \gg 1$ it can be brought to the form
written as
\be
G_{c~(2)}(r ,r';t) = {M \over 2 \pi} \int_{-m}^{m}~dw~e^{i (2 \pi \kappa - wt)}~e^{i \phi},
\ee
where we have defined
\be
e^{i \phi} = { 1 + (-1)^{2 \tim} e^{- 2 \pi i \kappa} \over
 1 + (-1)^{2 \tim} e^{2 \pi i \kappa}},
\ee
while $M$ provides the relation as follows:
\be
M = {(\Gamma(1 + 2\tim))^2~\Gamma(- 2\tim) \over 2 \tim ~\Gamma(2 \tim) }~(r r')^{1 \over 2}
\bigg[
J_{2 \tim} (\sqrt{8 \kappa \tom r})~J_{2 \tim} (\sqrt{8 \kappa \tom r'})
+ I_{2 \tim} (\sqrt{8 \kappa \tom r})~I_{2 \tim} (\sqrt{8 \kappa \tom r'}) 
\bigg].
\ee
At very late time both terms $e^{i w t}$ and $e^{2 \pi \kappa}$ are rapidly
oscillating. It means that the scalar waves are mixed states consisting of the states 
with multipole phases backscattered by spacetime curvature. Most of them cancel
with each others which have the inverse phase. In such a case, one can find the value of 
$G_{(2)}$ by means of the saddle point method. It could be found that the value $2 \pi \kappa
- wt$ is stationary at the value of $w$ equal to the following:
\be
a_{0} = \bigg[ { \pi~m (r_{+} + \alpha_{1}r_{-}) \over 2 \sqrt{2}} \bigg]^{1 \over 3}.
\ee
Then approximating integration by the contribution from the very close nearby of $a_{0}$
is given by
\be
F = {i m^{4/3} \over \sqrt{2}} (\pi)^{5 \over 6}(r_{+} + \alpha_{1}r_{-})^{1 \over 3}
(mt)^{-{ 5 \over 6}}~e^{i \phi(a_{0})}.
\ee
In comparison to the late-time behaviour of the second term in Eq.(\ref{fer}), the first term 
can be neglected. The dominant role plays the behaviour of the second term, i.e., the late-time 
behaviour is proportional to
${- {5 \over 6}}$. Thus, the asymptotic late-time behaviour of the Green's function 
can be written in the form 
\be            
G_{c}(r,r';t) = {m^{4/3} \over \sqrt{2}} (\pi)^{5 \over 6}(r_{+} + \alpha_{1}r_{-})^{1 \over 3}
(mt)^{-{ 5 \over 6}}~\sin(mt)~\tpsi(r, m)~\tpsi(r', m),
\ee

\section{Conclusions}
            
We have studied analytically the intermediate and the late-time behaviour of massive
scalar field in the background of a dilaton black hole arising in the low energy string theory, the
so-called dilaton gravity. We took into account the coupling constant $\alpha$ appearing in the 
action in dilaton gravity. In the case of the intermediate asymptotic behaviour we confirmed analytically
the results found in \cite{mod01b}, i.e., the oscillatory power law dependence only on the 
angulat momentum parameter $l$ and the field parameter, the mass $m$ of it. However, as was 
claimed in Ref.\cite{mod01b} the intermediate asymptotic 
behaviour is not the final pattern of decay rate. The decay rate of the form
$t^{-{ 5 \over 6}}$ occurs when $\kappa \gg 1$. It stems from th resonance backscattering
off the spacetime curvature. This tail rate behaviour is not dependent on
mass of the black hole and scalar field, $\alpha$ coupling, and multiple number $l$.           
Our analytical results confirmed the numerical predictions \cite{mod01b}
that oscillatory tail decay rate of the form $t^{-{ 5 \over 6}}$.
Our conclusions are also applicable for the extremal dilaton
black holes, i.e., for which one has $r_{+} = r_{-}$. 
Having in mind relation (\ref{ext}) one gets the following 
late-time behaviour for the extremal dilaton black hole. It provides the following:
\be
G_{c}^{ext}(r, r';t) = {m^{4/3} \over \sqrt{2}} (\pi)^{5 \over 6}~
\bigg( {2 Q \over \sqrt{1 + \alpha^2}} \bigg)^{1 \over 3}
(mt)^{-{ 5 \over 6}}~\sin(mt)~\tpsi(r, m)~\tpsi(r', m),
\ee





\end{document}